\documentclass[pra,twocolumn,showpacs,letterpaper,showpacs,superscriptaddress]{revtex4}

\usepackage{graphicx,amsmath,amssymb,amsfonts,latexsym,color,dcolumn,bm,epsfig,subfigure}

\begin{document}

\title{Surface Contact Potential Patches and Casimir Force Measurements}

\author{W. J. Kim$^*$}
\affiliation{Yale University, Department of Physics, P.O. Box 208120, New Haven, CT 06520-8120, USA}

\author{A.O. Sushkov}
\affiliation{Yale University, Department of Physics, P.O. Box 208120, New Haven, CT 06520-8120, USA}

\author{D. A. R. Dalvit}
\affiliation{Theoretical Division MS B213, Los Alamos National Laboratory, Los Alamos, NM 87545, USA}

\author{S. K. Lamoreaux}
\affiliation{Yale University, Department of Physics, P.O. Box 208120, New Haven, CT 06520-8120, USA}

\date{\today}

\begin{abstract}

We present calculations of contact potential surface patch effects that simplify previous treatments. It is shown that, because of the linearity of Laplace's equation, the presence of patch potentials does not affect an electrostatic calibration of a two-plate Casimir measurement apparatus. Using models that include long-range variations in the contact potential across the plate surfaces, a number of experimental observations can be reproduced and explained. For these models, numerical calculations show that if a voltage is applied between the plates which minimizes the force, a residual electrostatic force persists, and that the minimizing potential varies with distance.  The residual force can be described by a fit to a simple two-parameter function involving the minimizing potential and its variation with distance. We show the origin of this residual force by use of a simple parallel capacitor model.  Finally, the implications of a residual force that varies in a manner different from $1/d$ on the accuracy of previous Casimir measurements is discussed.

\end{abstract}


\maketitle


\section{Introduction}

It is often assumed that the surface of a conductor is an equipotential. While this would be true for a perfectly clean surface of a homogeneous conductor cut along one of its crystalline planes, it is not the case for any real surface. Potential patches can be caused by, for example, oxide films or some other films adsorbed on the surface (less than a monolayer is required), strains in the surface, and chemical impurities within the surface. Such patches effectively create a surface dipole layer, which alters the potential above the surface. Various types of monopolar charge disorder could be also present, significantly altering the properties of a surface from that of an ideal conductor  \cite{Rudi}. Even for chemically unreactive noble metals, such as gold and copper, carefully prepared in an ultra-clean environment to minimize such ``dirt'' films,  experiments show that typical surface potential variations are on the order of at least a few millivolts~\cite{Rose,Riviere,Michaelson}. Similar effects were found in a more recent measurement on a pair of metallic plates employed in the Laser Interferometer Gravitational Wave Observatory (LIGO) project \cite{LIGO}. The cause of surface potential variations is most likely local changes in surface crystalline structure, giving rise to varying work functions and hence varying-potential patches. It is well known that the work function of a metal surface depends on the crystallographic plane along which it lies; as an example, for gold the work functions are 5.47~eV, 5.37~eV, and 5.31~eV for surfaces in the $\langle 100 \rangle$, $\langle 110 \rangle$, and $\langle 111 \rangle$ directions, respectively \cite{CRC}. This variation is most likely due to the different effective electron masses, and the resulting different Fermi energies along the corresponding directions.

When two conductors of different work functions are brought into contact, electrons flow until the chemical potential (i.e., the Fermi energy) in both conductors equalizes. As a result, there is a net dipole distribution created at the interface, which gives rise to a ``contact'' or ``Volta'' potential established between the two conductors, equal to the difference between their work functions. We therefore expect that the electrostatic potential along a chemically clean metal surface varies on the length scale of the typical size of surface crystallites, which can vary from sub-micron to millimeter or larger scales.

The generation of an attractive force between conducting surfaces due to metallic contacts has long been recognized as a possible systematic limitation to a Casimir force measurement \cite{Sparnaay,StevePRL}. Recently, Speake and Trenkel \cite{Speake2003} have performed a formal treatment of the effect of random, zero-average, patch potentials. We reproduce here their result using a slightly different formalism, and show that the result can be simply obtained, expressed in terms of the surface potential autocorrelation function. Furthermore, we provide a simple model that shows the origin of a distance-dependent minimizing potential, and that this dependence leads to a residual electrostatic force that can have a complicated character. The goal of this paper is not to present a full rigorous mathematical derivation of these effects, but to illustrate their fundamental nature and motivate functional forms that have been observed in experiments in a straightforward manner. More importantly, if the patch potential is actually measured, it would be possible to perform an exact numerical calculation of the excess force. Our recent work \cite{ourwork2009} indicates variations of contact potentials on the level of 10 mV, which is a challenging level for Kelvin probe techniques with the appropriate spatial resolution \cite{kelvinprobe}. Thus, the full experimental description of these effects remains an open challenge.

In this paper we define a positive force as an attraction between the surfaces under consideration. Therefore, attractive electrostatic and Casimir forces are taken as positive.


\section{Attractive Electrostatic Force due to Small Surface Patches}

In this section we will consider the effect of random surface patches on the electrostatic interaction
between metallic plates, for the plane-plane and plane-sphere geometries. We will assume that the
typical patch area is much smaller than the effective area of the surface interactions, defined for each
of the geometries below.

\subsection{Parallel planes geometry}

Consider two plane parallel metallic surfaces at $z=0$ and $z=d$.  Let
the electrostatic potential at $z=0$ be $V(x,y,z=0)=V_a(x,y)$ and at $z=d$ be
$V(x,y,z=d)=V_b(x,y)$. The solution of Laplace's equation in the region $0<z<d$ can
be easily found separating variables in cartesian coordinates,  $V(x,y,z)=X(x) Y(y) Z(z)$,
where
\begin{equation}
\frac{1}{X} \frac{d^2 X}{dx^2} = - \alpha^2 ; \;\;
\frac{1}{Y} \frac{d^2 Y}{dy^2} = - \beta^2;  \;\;
\frac{1}{Z} \frac{d^2 Z}{dz^2} = \gamma^2,
\end{equation}
and $\gamma^2=\alpha^2+\beta^2$.
The general solution of Laplace's equation in this geometry
can be written as
\begin{eqnarray}
&& V(x,y,z)=\int_{-\infty}^{\infty}  d\alpha \;  d\beta
\left[ \cos(\alpha x) + A_{\alpha} \sin(\alpha x) \right]  \nonumber \\
&& \times \left[ \cos(\beta y) + A_{\beta} \sin(\beta y) \right]
\left[ B^{+}_{\gamma} e^{\gamma z} + B^{-}_{\gamma} e^{-\gamma z} \right] .
\end{eqnarray}
It is convenient to expand the boundary conditions
in a cosine Fourier series:
\begin{equation}
V_a(x,y)=\int \frac{d^2{\bf k} }{(2\pi)^2} V_{a,{\bf k}} \cos(k_x x) \cos(k_y y) ,
\end{equation}
and similarly for $V_b(x,y)$. Imposing the boundary conditions
we get $A_{\alpha}=A_{\beta}=0$, $\alpha=k_x$, $\beta=k_y$ (hence $\gamma=\sqrt{k_x^2+k_y^2}$),
$B^{+}_{\gamma} = (V_{b,\gamma} - V_{a,\gamma} e^{-\gamma d})/ 2 \sinh(\gamma d)$ and
$B^{-}_{\gamma} = (V_{a,\gamma} e^{\gamma d}  - V_{b,\gamma} )/ 2 \sinh(\gamma d)$.
Therefore, the general solution of Laplace's equation with the given boundary conditions on each plate is
\begin{eqnarray}
&& V(x,y,z) = \int \frac{d^2 {\bf k}}{(2 \pi)^2} \frac{\cos(k_x x) \cos(k_y y)}{2 \sinh(\gamma d)} \\
&& \times \left[ e^{\gamma z} \left( V_{b,{\bf k}} - V_{a,{\bf k}} e^{-\gamma d} \right) +
e^{- \gamma z} \left( V_{a,{\bf k}} e^{\gamma d} - V_{b,{\bf k}} \right) \right] \nonumber .
\end{eqnarray}

Now we calculate the electrostatic energy between the plates. The electrostatic energy density is
$u=\frac{\epsilon_0}{2} | {\bf E} |^2$, so the total energy $U_{pp}$ per unit area $A=L_x L_y$ is
\begin{eqnarray}
U_{pp} &=& \frac{\epsilon_0}{2} \frac{1}{A} \int_0^{L_x} dx  \int_0^{L_y} dy \int_0^d dz  \nonumber \\
&\times &  \left[ \left( \frac{\partial V}{\partial x} \right)^2 +
\left( \frac{\partial V}{\partial y} \right)^2 +
\left( \frac{\partial V}{\partial z} \right)^2
\right] .
\end{eqnarray}
In order to proceed, we will assume that the electrostatic patches are stochastic, uncorrelated between the different plates, and for a given plane the 2-point correlation function is diagonal in the
${\bf k}$-basis. That is
\begin{eqnarray}
&& \langle V_{a, {\bf k}} \rangle = \langle V_{b, {\bf k}} \rangle =
\langle V_{b,{\bf k}} V_{a,{\bf k}'} \rangle = 0 ;  \nonumber \\
&& \langle V_{a,{\bf k}} V_{a,{\bf k}'} \rangle = C_{a,{\bf k}} \; \delta^2({\bf k} - {\bf k}') ;  \nonumber \\
&& \langle V_{b,{\bf k}} V_{b,{\bf k}'} \rangle = C_{b,{\bf k}} \; \delta^2({\bf k} - {\bf k}') ,
\end{eqnarray}
where $\langle \ldots \rangle$ means stochastic average. Note that, as in \cite{Speake2003}, we assume zero
cross-correlation between the plates, and that the correlation function of each plate is independent of the
position of the other plate (i.e., independent of the distance $d$).

It is then easy to calculate each of the terms
in the expression for $U_{pp}$:
\begin{eqnarray}
&& \langle \left( \frac{\partial V}{\partial x} \right)^2 \rangle =
\int \frac{d^2 {\bf k}}{(2\pi)^2}
\frac{k_x^2 \sin^2(k_x x) \cos^2(k_y y)}{4 \sinh^2(\gamma d) } \nonumber \\
&& \times
\left\{
2 C_{b,{\bf k}} [ \cosh(2 \gamma z) - 1] + 2 C_{a,{\bf k}} [\cosh(2 \gamma(z-d)) -1]
\right\} ; \nonumber
\end{eqnarray}
\begin{eqnarray}
&& \langle \left( \frac{\partial V}{\partial y} \right)^2 \rangle =
\int \frac{d^2 {\bf k}}{(2\pi)^2}
\frac{k_y^2 \cos^2(k_x x) \sin^2(k_y y)}{4 \sinh^2(\gamma d) } \nonumber \\
&& \times \left\{
2 C_{b,{\bf k}} [ \cosh(2 \gamma z) - 1] + 2 C_{a,{\bf k}} [\cosh(2 \gamma(z-d)) -1]
\right\} ; \nonumber
\end{eqnarray}
\begin{eqnarray}
&& \langle \left( \frac{\partial V}{\partial z} \right)^2 \rangle =
\int \frac{d^2 {\bf k}}{(2\pi)^2}
\frac{\gamma^2 \cos^2(k_x x) \cos^2(k_y y)}{4 \sinh^2(\gamma d) } \nonumber \\
&&  \times
\left\{
2 C_{b,{\bf k}} [ \cosh(2 \gamma z) + 1] + 2 C_{a,{\bf k}} [\cosh(2 \gamma(z-d)) +1]
\right\} . \nonumber
\end{eqnarray}
The $x$ and $y$ integrations are trivial, since $L_x^{-1} \int_0^{L_x} dx \sin^2(k_x x) = L_x^{-1} \int_0^{L_x} dx \cos^2(k_x x) = 1/2$ and the same for $y$. For the $z$ integration we use
$\int_0^d dz [ \cosh(2 \gamma z) \pm 1 ] =\sinh(2 \gamma d)/2 \gamma \pm d$ and
$\int_0^d dz [ \cosh(2 \gamma (z-d)) \pm 1 ] =\sinh(2 \gamma d)/2 \gamma \pm d$.
Finally the electrostatic parallel-plate energy is
\begin{equation}
U_{pp} = \frac{\epsilon_0}{16} \int \frac{d^2{\bf k}}{(2\pi)^2}  \frac{\gamma \sinh(2 \gamma d)}{\sinh^2(\gamma d)} [C_{a,{\bf k}} + C_{b,{\bf k}} ] .
\end{equation}
In the special case of an {\it isotropic} patch distribution, the correlation functions depend only on
$k = | {\bf k} |$, that is $C_{a,{\bf k}} = C_{a,k}$ and $C_{b,{\bf k}} = C_{b,k}$. Then
\begin{equation}
U_{pp} = \frac{\epsilon_0}{16}  \frac{1}{2 \pi} \int_0^{\infty}  dk \frac{k^2 \sinh(2 k d)}{\sinh^2(k d)} [C_{a,k} + C_{b,k} ] .
\label{selfenergy}
\end{equation}
For the  plane-plane geometry, the effective area of interaction $A_{\rm eff}$ is the same as the total
area of the plate $A_{\rm eff} = A$. In this case, the small surface patch limit corresponds to $k^2 A  \gg 1$,
which basically means that we neglect finite-size effects in the computation of the electrostatic energy.

Let us analyze different limiting cases of Eq.(\ref{selfenergy}). For surface potential patches small with respect
to the effective interaction area ($k^2 A  \gg 1$) but large with respect to the plates separation ($k d \ll 1$ or
$d \ll \lambda$, where $\lambda$ is a characteristic length of a potential patch)
the energy scales as $1/d$,
which is the same as in the usual electrostatic case for fixed (non-stochastic) potential difference.
This limit $k d \rightarrow 0$ is essentially the ``proximity force approximation" (PFA), to be discussed more fully in the next Section, applied to the electrostatic problem with in-plane potential variations. In this approximation the net energy and force
is calculated by considering the attraction between paired infinitesimal surface elements on each plate,
and for large surface patches, the energy is then the additive sum of the usual $1/d$ plane-plane energies. Indeed, if we rewrite Eq. (8) defining
the root mean square (rms) potential fluctuations $V_{\rm rms}^2$ as
\begin{equation}
V_{\rm rms}^2=\frac{1}{8 \pi}\int_0^\infty dk\ k (C_{a,k}+C_{b,k}) \equiv \int_0^\infty dk\ k S(k) ,
\end{equation}
one obtains in the limit $k d\rightarrow 0$,
\begin{equation}
U_{pp}=\frac{\epsilon_0 V_{\rm rms}^2}{2 d}.
\end{equation}

In the opposite limit  (surface potential patches small with respect to the effective interaction area, $k^2 A  \gg 1$, and small
with respect to the plates separation, $k d \gg 1$) Eq.(\ref{selfenergy}) has an asymptotic behavior independent of the distance $d$. This  is an artifact of the
calculational method, that has included the self-energy of each plate. Following \cite{Speake2003}, we remove from the above the potential energy at infinite separation, in order to have an expression for the {\it interaction} energy only. Using that
$\sinh(2 k d) = 2 \sinh(k d) \cosh(k d)$, we get
\begin{eqnarray}
U_{pp} & = & \frac{\epsilon_0}{4} \int_0^{\infty} dk\ k^2 \left(\frac{2\cosh(k d)}{\sinh(k d)}-2\right) S(k) \nonumber \\
& = & \frac{\epsilon_0}{2} \int_0^{\infty}  dk \frac{k^2 e^{-k d}  }{\sinh(k d)} S(k).
\end{eqnarray}
Therefore, in the limit $k d \rightarrow \infty$ the interaction energy vanishes exponentially. The
reason is that in this case the patches are so small and change sign so rapidly that there is no net electric field at a large distance $d$ from a given plate, hence there is no interaction with the other plate.

Some remarks are in order. First,  we have assumed that the boundary conditions on the two plates
had only stochastic components fluctuating around 0. When there is an external fixed potential difference $V$ between the two plates, the energy is the sum of the usual $V^2$ term plus the
patch component calculated above (basically this is due to the linearity of Laplace's equation).
The cross-terms $V  V_{1,{\bf k}}$ and $V V_{2,{\bf k}}$ cancel upon taking stochastic average.
Second, when the cross-correlations $\langle V_{1,{\bf k}} V_{2, {\bf k'}} \rangle $ are not zero,
there is a possibility that the interaction energy depends on the relative lateral position of the two plates, and
hence it is possible to have a  {\it lateral} force between the plates due to stochastic patches.


\subsection{Sphere-plane geometry}

In order to compute the patch effect on the force in the sphere-plane configuration we make use of the
proximity force approximation. Just as in the case of roughness in Casimir physics,
one must distinguish between two PFAs: one is for the treatment of the curvature of the sphere (valid when
$d \ll R$, where $R$ is the radius of curvature), and the other one is the PFA applied to the surface patch distribution (valid when $k d \ll 1$). We assume that we are in the conditions for PFA for the curvature, but we keep $k d$ arbitrary.

For the sphere-plane geometry we define an effective area of interaction by calculating along the plane
surface the distance $r$ from the point of closest approach
($r=0$, corresponding to the minimal sphere-plane separation $d$) where the separation between the surfaces doubles.
For a given $r$, this latter separation is given by $d(r)=d+ R(1-\cos\theta)$, with
$\sin\theta=r/R$. In the limit $\theta \ll 1$ (which corresponds to
$R \gg d(r)$) we obtain
$d(r)=d + r^2 /  2R$. The condition for the surface separation to be double that of the closest-approach distance is $d(r_{\rm eff})=2 d = d + r_{\rm eff}^2 /2R$, and the effective area is then
\begin{equation}
A_{\rm eff}=\pi r_{\rm eff}^2=2 \pi R d.
\end{equation}
For a sphere of radius $R=15$ cm separated by a plane by a distance $d=1$ $\mu$m, the effective distance above defined is  $r_{\rm eff}=0.05$ cm and the effective area is $A_{\rm eff} = 0.009 \; {\rm cm}^2$. The small patch limit corresponds to surface patches
of area much smaller than this effective area  $A_{\rm eff}$, that is $k^2 A_{\rm eff}  \gg 1$.

In the proximity force approximation, the electrostatic force in the sphere-plane case
is $F_{sp}(d)= 2 \pi R U_{pp}(d)$, namely
\begin{equation}
F_{sp} =  \pi \epsilon_0 R \int_0^{\infty} dk \frac{k^2 e^{-k d}  }{\sinh(k d)} S(k).
\label{forcesp}
\end{equation}
There are a number of models that can be used to describe the surface fluctuations.  The simplest is to say that the potential autocorrelation function is, for a distance $r$ along a plate surface,
\begin{equation}
{\cal R}(r) = V_0^2e^{-r^2/\lambda^2} .
\end{equation}
Then, by the Wiener-Khinchin theorem, the power spectral density $S(k)$ can be evaluated as the cosine two-dimensional Fourier transform of the autocorrelation function, which in our notation is \cite{stein}
\begin{equation}
S(k)=2\cdot {V_0^2}{\lambda^2\over 2} e^{-\pi\lambda^2k^2},
\end{equation}
where the factor of two reflects the statistically-independent contributions from the two plate surfaces ($C_{1,k}+C_{2,k}$).
The plane-sphere force is then given by, using  $k=u/\lambda$,
\begin{equation}
F_{sp}  = 2\pi \epsilon_0 R {V_0^2\over \lambda}\int_0^\infty du\ u^2 {e^{-\pi u^2} \over e^{2ud/\lambda}-1} .
\end{equation}
The limit of small potential patches $k d \rightarrow 0$ (and also small with respect to the effective area of interaction,
$k^2 A_{\rm eff}  \gg 1$), the force is
\begin{equation}
F_{sp}\approx {\epsilon_0 R V_{0}^2\over d} ,
\end{equation}
suggesting that $V^2_{\rm rms}=V_0^2/\pi$.  For the large $k d$ limit, the force becomes exponentially small,
just as in the parallel-plates geometry.

Another possible model for the patch distribution is the one used in \cite{Speake2003}.
Assuming that
$C_{a,k}=C_{b,k}={\tilde V}^2_0={\rm const}$ for $k_{\rm min} < k < k_{\rm max}$ and zero otherwise,
we get an expression for $V_{\rm rms}^2$ (similar to Eq. (12) in \cite{Speake2003})
\begin{eqnarray}
V_{\rm rms}^2
&=& \frac{1}{8 \pi}\int_0^\infty dk  k (C_{a,k}+C_{b,k}) =
\frac{{\tilde V}_0^2}{8 \pi}  (k_{\rm max}^2-k_{\rm min}^2)  \nonumber \\
&=& \int_0^{\infty} dk k S(k) .
\end{eqnarray}
We then obtain
\begin{equation}
S(k)=\frac{{\tilde V}_0^2}{4\pi}= \frac{2V_{\rm rms}^2}{k_{\rm max}^2-k_{\rm min}^2},
\end{equation}
for $k_{\rm min} < k < k_{\rm max}$ and zero otherwise.
The sphere-plane force from Eq. (\ref{forcesp}) therefore is
\begin{equation}
F_{sp} = \frac{4 \pi \epsilon_0 V_{\rm rms}^2 R}{k^2_{\rm max} - k^2_{\rm min}}
\int_{k_{\rm min}}^{k_{\rm max}} dk \frac{ k^2 e^{-k d}}{\sinh(k d)} ,
\end{equation}
which is the identical (apart from an overall, conventional sign) to Eq. (14) of \cite{Speake2003}.


\section{Attractive force due to large surface patches}

When the surface patches are larger than the effective area of interaction $A_{\rm eff}$ defined in the previous section, the force
between the plane and the sphere due to electrostatic patches can still be calculated using the method described above, but the average potential, as inferred by measuring the voltage at which a minimum in the attractive electrostatic force occurs, will vary with distance. This can be thought of as a finite-size effect; if the patch size is roughly the diameter of the plates, then there can be a non-zero average over the surface. Alternatively, this problem can be addressed by assuming a slowly varying average potential across the plate surfaces, as developed below.

In the PFA, the plate surfaces are divided into differential areas, and the attractive force, given the potential difference between the plates and the derivative of the capacitance between them, is calculated. Specifically, for the sphere-plate geometry,
\begin{equation}
F(d)=  \frac{\epsilon_0}{2} \int_0^{2 \pi} d\varphi   \int_0^R r dr
\frac{V^2(r,\varphi)}{(d+r^2/2R)^2} ,
\end{equation}
where $V(r,\varphi)$ is the net potential difference between the surface differentials, located at
$(r,\varphi)$ relative to the point of closest approach ($r=0$).  Writing $V$ in this form allows the possibility that there can be a slow (coherent) variation across the surface, as opposed to, and in addition to, innumerable small random patches. The PFA works here for two reasons.  One is that the lines of electric force do not cross each other,  the second is that the radius of curvature is large, so the angular deviations of the field lines are small when the plate diameter $D$ satisfies $D \ll R$. In this limit, the assumption that each surface differential element interacts only with a single element in the other plate is a good approximation.

A slow variation in potential across the plate surfaces will manifest itself as a distance variation in the potential that minimizes the electrostatic attractive force, i.e., $V_m=V_m(d)$. Specifically, if we define the force with some externally applied voltage $V_0$ to be
\begin{equation}
\label{force}
F(d,V_0)= \frac{\epsilon_0}{2}
\int_0^{2 \pi} d\varphi   \int_0^R r dr
\frac{(V(r,\varphi)+V_0)^2}{(d+r^2/2R)^2} ,
\end{equation}
the minimized force at a fixed distance determines the minimizing potential,
\begin{equation}
0= \left. \frac{ \partial F(d,V_0)}{\partial V_0} \right|_{V_0=V_m}=
\epsilon_0 \int_0^{2 \pi} d\varphi   \int_0^R r dr
\frac{V(r,\varphi)+V_m}{(d+r^2/2R)^2} .
\nonumber
\end{equation}
This equation implies a minimizing potential dependent on distance, $V_m=V_m(d)$.
Note that, in the idealized case of an equipotential
surface, i.e., $V(r,\varphi)={\rm const}$, $V_m$ would be independent of $d$, and the minimized electrostatic force $F(d,V_0=V_m)$ vanishes.
Incidentally, the second derivative of $F(d,V_0)$ with respect to $V_0$ can be used to determine the distance at which the measurement is being made,
\begin{equation}
\frac{\partial^2 F(d,V_0)}{\partial V_0^2}= 2 \pi \epsilon_0
\int_0^R dr \frac{r}{(d+r^2/2R)^2}
\approx \frac{2\pi R\epsilon_0}{d} ,
\nonumber
\end{equation}
where the finite size effects are neglected (upper limit of the $r$ integration is set to infinity, which is a very good approximation when $d \ll R$).  The important implication is that the patch potentials do not interfere with the electrostatic calibration, that is the fundamental basis of our experiment \cite{ourwork2009}, and of all Casimir force experiments.

It is worth emphasizing a couple of points. First, that the origin of the distance dependence of the minimizing potential $V_m(d)$ is an interplay between the curvature of the surfaces and a variation of the electrostatic potentials $V_i$ ($i=a,b$) along the surfaces (possibly due large surface patches).  We have shown above how this effect arises in the context of the sphere-plane geometry, but of course it can be easily generalized to any geometry involving non-planar surfaces. It also follows from the above that for the parallel plates geometry one should expect that, even
when the surface potentials $V_i=V_i(x,y)$ vary along the plane surfaces, there should be no interplay with the (infinite) curvature of the
planes, and thus the minimizing potential $V_m$ should be distance-independent \cite{ruoso2009}. 
Second, distance dependence of the electrical potential minimizing the force between the plates has been observed in a number of experiments  in the sphere-plane geometry\cite{kim2008,iannuzzi2009,pollack}, as well as in our own work \cite{ourwork2009}, with further investigations under way.


\begin{figure}[t]
\begin{center}
\includegraphics[width=0.8\columnwidth,clip]{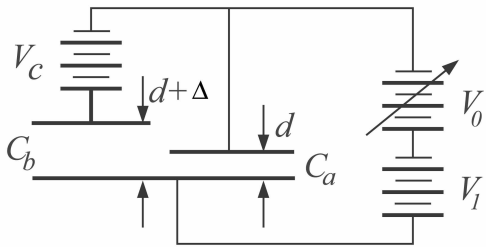}
\caption{A toy model illustrating the mechanism for the generation of a distance-dependent minimizing electrostatic potential $V_m(d)$ and electrostatic residual force $F^{\rm el}_{\rm res}(d)$.}
\label{pmod}
\end{center}
\end{figure}

It has been suggested that the variation of the minimizing potential with distance can cause an additional electrostatic force
$F(d,V_0=V_m(d))$, and an estimate was made for the possible size of the effect \cite{myarxiv}, where the varying contact potential is considered in a system of plates connected in series. The analysis presented in \cite{myarxiv} does not reproduce the effects seen in our experimental work, and we were unable to develop a fundamental theory of a plate-plate interaction that could cause a varying contact potential. Nonetheless, we have seen above that within the PFA a coherent variation of the surface potential along
the non-planar surface does imply a distance-dependent minimizing potential. 
Now we present a simple model that produces not only varying contact potentials, but also the corresponding residual electrostatic force, consistent with our observations in \cite{ourwork2009}. The model is depicted in Fig. \ref{pmod}. In this figure, the two capacitors (short distance, $C_a(d)$, long distance, $C_b(d+\Delta)$) create a net force on the lower continuous plate (setting $V_1=0$ initially),
\begin{equation}
F(d,V_0)=-\frac{1}{2} C_a' V_0^2 - \frac{1}{2} C_b'(V_0+V_c)^2,
\end{equation}
where
\begin{equation}
C_a'= {\partial C_a(d)\over \partial d};\ \ \  C_b'={\partial C_b(d+\Delta)\over \partial d} ,
\end{equation}
and $V_0$ can be varied, with $V_c$ a fixed property of the plates. The force is minimized when
\begin{eqnarray}
\left. {\partial F(d,V_0)\over \partial V_0} \right|_{V_0=V_m} = 0
& \Rightarrow & V_m(d)=-{C_b' V_c \over C_a'+C_b'} ,
\end{eqnarray}
implying a residual electrostatic force
\begin{eqnarray}
F^{\rm el}_{\rm res}(d) &=& F(d,V_0=V_m(d)) \nonumber \\
& =& -\left[C_a'+{C_a'^2\over C_b'}\right]{V_m^2(d)\over 2} \nonumber \\
&=& -\left[ \frac{C_a' C_b'}{C_a'+C_b'} \right]{V_c^2\over 2} .
\end{eqnarray}
It is easy to take a case of parallel plate capacitors
($C_a'=-\epsilon_0 A/d^2$ and $C_b'=-\epsilon_0 A/(d+\Delta)^2$, where
$A$ is the area of each of the upper plates in Fig. 1, assumed to be equal;
hence, the lower continuous plate has area $2 A$)
and to show that there is a residual electrostatic force at the minimizing potential. Indeed, in such case,
\begin{eqnarray}
V_m(d) &=& - V_c \frac{d^2}{d^2 + (d+\Delta)^2} ,
\label{toyvm}\\
F^{\rm el}_{\rm res}(d) &=& \frac{\epsilon_0 A}{2} \; \frac{V_c^2}{d^2 + (d+\Delta)^2} .
\end{eqnarray}
Alternatively, in terms of $V_m(d)$ (up to $V_1$, see below), the force is
\begin{equation}
\label{fres}
F^{\rm el}_{\rm res}(d) = \frac{\epsilon_0 A}{2} \; \frac{V_m^2(d)[d^2 + (d+\Delta)^2]}{d^4} .
\end{equation}
Experimentally, $V_m(d)$ must include a distance-independent offset $V_1$ which arbitrarily depends on the sum of contact potentials in the complete circuit between the plates.
Therefore, the force due to large patches considered in this section should be written as proportional to $(V_m(d)+V_1)^2$, instead of simply $V_m^2(d)$, where $V_1$ is determined by a fit to experimental data. Note that when the two upper capacitor plates are at the same distance from the lower capacitor plate, i.e.
 when $\Delta=0$, Eq.(\ref{toyvm}) predicts a $V_m$ independent of distance, and Eq.(\ref{fres}) predicts a residual electrostatic
 force $F^{\rm el}_{\rm res}(d)  \propto (V_m + V_1)^2/d^2$. Since no residual force is expected in this case, the minimizing
 potential must be $V_m=-V_1$.

In order to analyze the sphere-plane geometry, one can divide the sphere into infinitesimal planar areas (as done
in the proximity force approximation), each with a random potential. In this picture, one can think of
the two upper capacitor plates in Fig. 1 as one of those infinitesimal parts of the
whole spherical surface, and the distance $\Delta$ being a local distance $\Delta(r)=d+R (1-\cos \theta)$ reflecting the effect of the
curvature of the spherical surface.  In this case, $C_a'(d)=-2\pi \epsilon_0 R/d$, and the denominator of Eq. (\ref{fres}) becomes $d^2$.
Integrating the force on the lower planar plate over the whole spherical surface to get the net force leads to a further reduction of the power of $d$ in the denominator, leaving the sphere-plane residual electrostatic force proportional to $(V_m(d)+V_1)^2/d$. Again, $V_1$ is a fit parameter that represents a sort of surface average potential, plus circuit offsets.


\section{Total Electrostatic Force Residuals}

We are now in a position to compute the total electrostatic residual force at the minimizing potential.  On the one hand, the presence
of surface patches small with respect to the effective area of interaction in the sphere-plane geometry leads to an attractive electrostatic
force given by Eq.(\ref{forcesp}), which was derived using the proximity force approximation ($d \ll R$) to treat the curvature
of the spherical plate. This component of the force due to stochastic potential patches is clearly independent of the applied
voltage $V_0$ between the plates, and therefore will be present even when the applied voltage is set at the minimizing potential,
$V_0=V_m(d)$. As we have seen in Section II, the exact dependence of this force on distance varies for different models
for the statistical properties of the two-point correlation functions of the surface potentials. However, in the limit of patches much
smaller that the sphere-plane separation ($k d \ll 1$), all models predict a dependency of the form $V_{\rm rms}^2/d$, where
$V_{\rm rms}$ denotes the rms voltage fluctuations.

On the other hand, the presence of surface patches large with respect to the effective area of interaction, corresponding to a coherent
variation of the potential along the spherical surface $V(r,\varphi)$, leads both to a distance-dependent minimizing potential
$V_m(d)$ and to a residual component of the electrostatic force $F^{\rm el}_{\rm res}(d) = F(d,V_0=V_m(d))$, even when the external
potential $V_0$ is set at the minimizing potential. As seen in Section III, this force is due to the interplay between surface patches and the curvature of the non-planar surface, has the form $(V_m(d) + V_1)^2/d$,  and is in addition to the $V_0$-independent force due to small
potential patches.

By adding the two contributions of small and large surface patches (Section II and III, respectively),
we can determine the form of the residual electrostatic force at the minimizing potential. By numerical modeling of different patch sizes and distributions using Eq. (\ref{force}), we found that the experimental observations should be described by a relationship of the form
\begin{equation}
F^{\rm el}_{\rm res}(d)= \pi R\epsilon_0 \left[\frac{ (V_m(d)+V_1)^2+V_{\rm rms}^2}{d}\right],
\label{totalres}
\end{equation}
where we have specifically assumed the small $k d$ limit for the small patches. Both $V_1$ and $V_{\rm rms}$ are constants determined by fitting to the observed force at large plate separations, where the Casimir force is relatively small.
In \cite{ourwork2009} we have measured the dependency of the minimizing potential with distance, and
we have applied Eq.(\ref{totalres}) to fit the observed residual electrostatic force in a Ge sphere-plane
Casimir experiment for distances large enough ($d > 5 \mu$m) to neglect possible contributions
due to Casimir forces. With only two adjustable parameters $V_1$ and $V_{\rm rms}$, very good fit ($\chi^2$ of order unity) between the experimental data and Eq.(\ref{totalres}) was possible. We emphasize that Eq. (\ref{totalres}) was obtained in a heuristic way by numerically analyzing the distance-dependent force and variation in minimizing potential due to random surface potential distributions by use of Eq. ({\ref{force}). The physical basis of Eq. (\ref{totalres}) is nonetheless quite sound.


\section{Distance correction due to residual forces and its implication on the interpretation of previous data}

In previous work \cite{StevePRL}, the minimizing potential was assumed to be independent of distance and the absolute separation between the sphere-plane plates in a Casimir experiment was determined by fitting a background electrostatic force (at distances where the Casimir force is relatively small and negligible) to a function
\begin{equation}
F^{\rm el}_{\rm res,fit}(d)={B_f \over d+d_{0f}},
\label{fitfunction}
\end{equation}
where $B_f$ and $d_{0f}$ are fitting constants. Unfortunately, in \cite{StevePRL}, a possible distance variation of the minimizing potential was not measured. However, a relatively large fixed potential was constantly applied to give a $1/d$ force in the data, and this was used to determine the absolute separation. 

One can estimate a systematic error for \cite{StevePRL} (and possibly for other experiments where $V_m$ was assumed to be constant) by considering a possible distance-dependence of the minimizing potential, and taking our Ge measurement results \cite{ourwork2009} as ``typical".
In the previous section we noted that the residual long range force in the Ge experiment is well described by Eq.(\ref{totalres}) in terms of the measured distance-dependent
contact potential $V_m(d)$. If, instead, we had mistakenly assumed $V_m$  to be independent of distance and neglected any $V_{\rm rms}$ contribution, the
same experimental data for the residual long range force in \cite{ourwork2009} is also well described by a power-law fit of the form $1/d^e$, with exponent $e=0.72 \pm 0.2$ (as mentioned in footnote [24] in \cite{ourwork2009}). Note that this is the residual force at the minimizing potential, so if the applied voltage is not set precisely to $V_m(d)$, there can be an additional force, but we neglect that here.  In the following we will  assume that a power-law form for the long-range residual force also applies to the Au experiment
\cite{StevePRL}, and we will take a nominal value of $0.8$ for the exponent to assess the possible systematic error introduced in \cite{StevePRL}.  This estimate is entirely heuristic, and given that the minimizing potential was not measured as a function of distance in  \cite{StevePRL}, it is the best that can be done in a post-analysis and represents a reasonable range.

An error in the distance determination is introduced if an improper function $1/d$ is used instead of the ``true'' function, which 
we take to be $1/d^{0.8}$. Generally, we might expect contributions from both, as
\begin{equation}
F^{\rm el}_{\rm res,true}(d)={B_1 \over d+d_{0}}+{B_2\over (d+d_0)^{0.8}},
\label{TRUEfunction}
\end{equation}
where the first term represents a patch potential force and an inexact cancelation of the average contact potential ($V_1$), and the second term represents the force due to the variation in $V_m(d)$. We assume that $d_0$ is fixed and known; our goal is to determine the distance error in determining $d_0$ by fitting to a function of the wrong form.
We can define an effective $\chi^2$ and minimize its value to find the offset due to the possibility that the wrong fitting function, $1/(d+d_0)$ alone, was used in \cite{StevePRL}.  A form that allows simple numerical calculation is as follows, substituting $x=d/d_0$
\begin{equation}
\chi^2=
\int_0^{x_{\rm max}} dx  \Bigg[ {B_f \over x+(1+\epsilon )}
-{\alpha\over (x+1)^{0.8}}
-{(1-\alpha) \over x+1}\Bigg]^2 ,
\label{minichi2}
\end{equation}
where $d_{0f}=d_0+\epsilon d_0$, $\alpha$ parameterizes the relative amounts, at $d=0$ (or $x=0$) of the $1/d$ and $1/d^{0.8}$ forces in the ``true" (and assumed known) function,
and $x_{\rm max}=d_{\max}/d_0$ is the largest (dimensionless) separation to which measurements are taken.
When $\alpha=0$ the ``wrong" fitting function is equal to the ``true" function, and $\epsilon=0$. For simplicity, we choose to define $\alpha$ as the relative contribution of the two forces at $d_0$ because their relative size is distance-dependent.

In \cite{StevePRL} the relative contributions from an applied fixed voltage $1/d$ force, and the possible residual unaccounted $1/d^{0.8}$ force implied by our recent results \cite{ourwork2009}, are in the ratio approximately 10/1, so $\alpha\approx 0.1$, at distances of order $1\ \mu$m.
The parameters $B_f$ and $\epsilon$ are then determined by minimizing $\chi^2$.
Although the integral in Eq.(\ref{minichi2}) appears as elementary, its evaluation is quite cumbersome.  Results of numerically minimizing $\chi^2$ as a function of $\alpha$ and $B_f$, with $x_{\rm max}=10$, show that
\begin{equation}
\epsilon=0.65\alpha .
\end{equation}
Therefore, for $\alpha=0.1$ we obtain $\epsilon=0.065$.  Applying this result to Fig. 4 in \cite{StevePRL} shows that the $x$-axis needs to be shifted to the left (toward the origin) by $0.065\ \mu$m. That is, the distance scale is offset by 0.065 $\mu$m.  With this displacement, the Casimir force $F_c$ at the true location is better described by the theoretical result due to B\"ostrom and Sernelius \cite{bands}, which can be seen from the fractional change in force, $ \delta F_c/F_c=-3\delta d/d$ , which gives a 20\% effect and brings the measured Casimir force into agreement with the predicted result in \cite{bands} for distances around 1 $\mu$m; however, this result should be considered as preliminary and as a rough estimate of the correction magnitude that is possible. Because patch potential effects are sample dependent, it is not possible to say conclusively that the effects described here contributed to the result, although, not having tested for such possible effects, an additional systematic error could be ascribed to the result in \cite{StevePRL}.

Other possible background forces that deviate from a $1/d$ character will lead to corrections to the true distance when a fit to $1/(d-d_0)$ is performed. For the discussion here, we chose a form that is motivated by our experimental results, and by a theoretical analysis \cite{Speake2003}.  Alternatively, if a residual force that appears to have a $1/d$ character is removed from Casimir data, a possible background force as considered here will persist as a direct systematic.


\section{Conclusions}

We have derived in a straightforward and heuristic manner several important results pertaining to the excess electric force between plates that results from random surface patch potentials. These results have been cast in terms of the surface autocorrelation function, or alternatively the two dimensional spatial Fourier power spectrum. Our recent measurement \cite{ourwork2009} of short-range forces with Ge plates in the sphere-plane configuration has shown the importance of assessing surface patch potentials when measuring Casimir force residuals.  A recent Casimir experiment in the plane-plane configuration \cite{ruoso2009} has also found a large residual (non-Casimir) force that is also probably related to electrostatic patch residual forces.

Furthermore, we have shown that long-range surface correlations in the sphere-plane geometry can lead to a distance dependence of the electrostatic force minimizing potential. This effect is due to the dependence of the net surface averaging area on the separation between the plates and has been described by use of a simple capacitor model. The model clearly reproduces the general effects, provides an explanation of the origin of the varying minimizing potential, and demonstrates that even when the force is minimized, a residual electrostatic force remains. The results here should be compared to earlier work \cite{myarxiv} where it was assumed that the variation in minimizing potential was due to a voltage in series with the plates, with that voltage varying with distance by some unknown mechanism; this analysis could not describe the observed (non-Casimir) force in the germanium measurements described in \cite{ourwork2009}. We note that the effect is expected to exist whenever there are surface potential patches and non-perfectly-parallel surfaces; the model shown in Fig. 1 requires only the existence of patches with different absolute distances between the surfaces.  Such distance differentials can be created by surface roughness and/or lack of parallelism. 

Finally, we have shown that determining the distance between a sphere and a plane in a Casimir force experiment can be subject to systematic effects arising from residual electrostatic forces. The magnitude of the error in the distance determination is large enough to bring the results presented in \cite{StevePRL} into agreement with the calculation that takes into account properly the low frequency permittivity of metals \cite{bands}. The relevance of the analysis presented in this paper to precision Casimir force measurements and their possible systematic contamination should not go unnoticed. We are currently performing new measurements using Au coated plates in the apparatus used for Ge measurements, and will revisit these systematic effects.

\vspace{0.1cm}
\noindent
$^*${Present address:  Dept. of Physics, Seattle University, 901 12th Avenue, Seattle, WA 98122}
\begin{acknowledgments}

S.K.L. and A.O.S.'s work was funded by Yale University, and DARPA/MTO's Casimir Effect Enhancement project under SPAWAR contract number N66001-09-1-2071.  D.A.R.D.'s work was funded by DARPA/MTO's Casimir Effect Enhancement project under DOE/NNSA Contract
DE-AC52-06NA25396. He is grateful to R. Onofrio for insightful discussions.

\end{acknowledgments}


\end{document}